\documentstyle[12pt,aaspp4]{article}
\newcommand{\beq}{\begin{equation}}
\newcommand{\beqa}{\begin{eqnarray}}
\newcommand{\eeq}{\end{equation}}
\newcommand{\eeqa}{\end{eqnarray}}
\newcommand{\simg}{\gtrsim}
\newcommand{\siml}{\lesssim}

\newcommand{\bmn}{{\hat {\bm n}}}
\newcommand{\bmd}{{\hat {\bm d}}}
\newcommand{\bmL}{{\hat {\bm L}}}
\newcommand{\bm}[1]{\mbox{{\boldmath $#1$}}} 
\def\threedots{\mbox{$\cdot$\kern-1ex$\cdot$\kern-1ex$\cdot$}}
\def\dddot#1{\mathop{#1}^{\threedots}}
\def\fourdots{\mbox{$\cdot$\kern-1ex$\cdot$\kern-1ex$\cdot$\kern-1ex$\cdot$}}

\def\bari{\hbox{$I$\kern -.51em {\raise
      .53ex\hbox{$\scriptscriptstyle  - $}}}}
\begin{document}
%
%{KUNS-1593,YITP-99-49 }
\title{ Gravitational waves from inspiralling compact binaries\\ with 
magnetic dipole moments }
\author{
Kunihito Ioka$^{1}$
and Keisuke Taniguchi$^{2}$
}
\affil{$^{1}$ Department of Physics, Kyoto University, Kyoto 606-8502,
Japan}
\affil{
$^{2}$Yukawa Institute for Theoretical Physics, Kyoto University, 
Kyoto 606-8502, Japan\footnote[3]{
Present address : D\'epartement d'Astrophysique Relativiste et de Cosmologie,
UPR 8629 du C.N.R.S., Obserbatoire de Paris, F-92195 Meudon Cedex, France
}}
\centerline{JAN~~7~~2000}
\authoremail{iokakuni@tap.scphys.kyoto-u.ac.jp}

\begin{abstract}

We investigate the effects of the magnetic dipole-dipole coupling and
the electromagnetic radiation on the frequency evolution of
gravitational waves from inspiralling binary neutron stars with
magnetic dipole moments. This study is motivated by the discovery of
the superstrongly magnetized neutron stars, i.e., magnetar. We derive
the contributions of the magnetic fields to the accumulated cycles in
gravitational waves as $N_{mag} \sim 6 \times 10^{-3} (H/10^{16}{\rm
G})^{2}$, where $H$ denotes the strength of the polar magnetic fields
of each neutron star in the binary system. It is found that the
effects of the magnetic fields will be negligible for the detection
and the parameter estimation of gravitational waves, if the upper
limit for magnetic fields of neutron stars are less than $\sim
10^{16}$G, which is the maximum magnetic field observed in the soft
gamma repeaters and the anomalous X-ray pulsars up to date. We also
discuss the implications of electromagnetic radiation from the
inspiralling binary neutron stars for the precursory X-ray emission
prior to the gamma ray burst observed by the Ginga satellite.

\end{abstract}


\section{INTRODUCTION}\label{sec:intro}

Direct detection of gravitational waves (GWs) is one of the most
exciting challenges in the history of science. Long baseline
interferometers for detection of GWs such as TAMA300 (Kuroda et
al. 1997), GEO600 (Hough 1992), VIRGO (Bradaschia et al. 1990), and
LIGO (Abramovici et al. 1992) will be in operation within five years. 
One of the most promising sources of GWs for such detectors
is the inspiralling binary neutron stars (BNSs), since we may expect
several coalescing events per year within 200Mpc (Phinney 1991,
Narayan, Piran, \& Shemi 1991, van den Heuvel \& Lorimer 1996).

As the orbital radius of BNSs decays due to gravitational radiation
reaction, the frequency of GWs sweeps upward in detector's sensitive
bandwidth from $\sim 10$Hz to $\sim 1000$Hz. In the early inspiralling
phase of BNSs, each neutron star (NS) can be treated as a point
particle and the post-Newtonian (PN) expansion will converge (Cutler
et al. 1993) since the orbital separation of NSs is much larger than
the NS's radius and the orbital velocity is much smaller than the
velocity of light. This means that the theoretical templates of GWs in
the inspiralling phase can be calculated with high accuracy by the PN
approximation of general relativity using only several parameters:
each mass and spin of NSs and the initial orbital elements. By cross
correlating the observed noisy signals with the theoretical templates,
the binary parameters, such as masses, spins (Cutler et al. 1993,
Kidder, Will, \& Wiseman 1993, Cutler \& Flanagan 1994, Poisson \&
Will 1995), and cosmological distances (Schutz 1986), are
determined. The number of rotation of BNSs is about 16000 in the
detector's sensitive bandwidth. Therefore, the quite accurate
theoretical templates are needed in order to extract physical
information about BNSs from GWs since any effect that causes only one
cycle ambiguity over 16000 accumulated cycles in the theoretical
templates will reduce the signal-to-noise ratio (SNR).
For the inspiralling BNSs in the sensitive bandwidth, with $v^2 \sim
m/r$ (hereafter $G=c=1$) typically around $10^{-2}$, the correction of
$1/16000 \sim 10^{-4}$ corresponds to second PN (2PN) order, $(m/r)^2
\sim 10^{-4}$ (Blanchet et al. 1995). Many efforts are devoted to
calculate higher-order PN corrections to theoretical templates
(e.g., Blanchet 1996, Jaranowski \& Sch$\ddot{\rm a}$fer 1998a, 1998b,
Damour, Jaranowski \& Sch$\ddot{\rm a}$fer 1999,
Tagoshi \& Nakamura 1994, Tagoshi \& Sasaki 1994, Poisson 1995). 
However, these studies pay attention only to the
gravitational effects on the theoretical templates, and it has not
been studied how large corrections to the theoretical templates are
caused by the electromagnetic effects, i.e., the magnetic fields of
NSs, as far as we know.\footnote{
Of course, the electromagnetic effects on the GWs from a rotating NS
are investigated vigorously (Bocquet et al. 1995, Bonazzola \&
Gourgoulhon 1996, Konno, Obata, \& Kojima 1999).
}

NSs observed as radio pulsars are believed to have strong magnetic
fields, typically $\sim 10^{12}$G, assuming that the spin-down of
pulsars is due to magnetic dipole radiation (e.g. Taylor, Manchester,
\& Lyne 1993). In addition, it begins to be recognized recently that
NSs with superstrong magnetic fields $\simg 10^{14}$G really exist
(see below). We can crudely estimate the correction to the waveform
due to magnetic dipole fields of NSs by comparing the gravitational
force $F_{G} \sim m_1 m_2 / r^2$ and the magnetic force $F_{M} \sim 3
\mu_1 \mu_2 / r^4$ between BNSs of masses, $m_1$ and $m_2$, with
magnetic dipole moments, $\mu_1$ and $\mu_2$, as
\beqa
{{F_M}\over{F_G}} &=& 1 \times 10^{-4}
\left({{r}\over{6 (m_1+m_2)}}\right)^{-2}
\left({{H_1}\over{2 \times 10^{16} {\rm G}}}\right)
\left({{H_2}\over{2 \times 10^{16} {\rm G}}}\right) \nonumber \\
&&\times
\left({R_1 \over 10^6 {\rm cm}}\right)^3
\left({R_2 \over 10^6 {\rm cm}}\right)^3
\left({1.4M_{\odot} \over m_1}\right)
\left({1.4M_{\odot} \over m_2}\right)
\left({2.8M_{\odot} \over m_1+m_2}\right)^2,
\label{eq:crude}
\eeqa
where $H_p=2\mu_p/R_p^3$ $(p=1,2)$ are the magnetic fields at the pole
of the star and $R_p$ are radii of NSs.  This corresponds to the 2PN
order correction. Therefore, the magnetic fields of order $\sim
10^{16}$G might cause about one rotation error.  Note that
the $r$-dependence of the magnetic correction is the same as that of 2PN
order, $v^4 \propto r^{-2}$, so that this argument is independent of
the value of the separation $r$.

Theoretically, in a new born NS, such superstrong magnetic fields
$\sim 10^{16}$ G can be generated if the initial spin is in millisecond
range since the conditions for helical dynamo action are met during
the first few seconds after gravitational collapse (Duncan \& Thompson
1992, Thompson \& Duncan 1993). Observationally, such a superstrongly
magnetized NS, or ``magnetar'', may be found as the soft gamma
repeaters (SGRs) and the anomalous X-ray pulsars (AXPs) (Duncan \&
Thompson 1992, Thompson \& Duncan 1996).
%SGRs are persistent X-ray sources with X-ray luminosity $\sim
%10^{35}$--$10^{36}$ ergs s$^{-1}$ that emit brief ($\sim 0.1$ s),
%intense ($\sim 10^{39}$--$10^{42}$) ergs s$^{-1}$ recurrent bursts of
%soft ($\sim 30$ keV) gamma ray (see Kouveliotou 1999 for a review).
%There are four known SGRs, SGR1900+14 (Mazets, Golenetskii, \& Guryan
%1979, Woods et al. 1999b), SGR1806-20 (Laros et al. 1986, Hurley et
%al. 1999), SGR1627-41 (Woods et al. 1999a, Mazets et al. 1999) in the
%Galactic plane and SGR0525-66 (Mazets et al. 1979, Marsden et
%al. 1996) in the Large Magellanic Cloud.
The dipolar magnetic fields of SGRs are estimated as $\sim 10^{15}$G
by using the measured spin periods with spin-down rates for SGR1900+14
(Kouveliotou et al. 1999, Woods et al. 1999b) and SGR1806-20
(Kouveliotou et al. 1998), and with the peak luminosity for SGR1627-41
(Woods et al. 1999a).
%AXPs are a group of six to eight pulsating X-ray sources with periods
%around 5--12 s, which are anomalous in comparison with average
%characteristics of known accreting X-ray pulsars (Mereghetti \& Stella
%1995, van Paradijs, Taam, \& van den Heuvel 1995); they are soft X-ray
%sources having low luminosities of $\sim 10^{34}$--$10^{35}$ ergs
%s$^{-1}$, show no sign of any companion, are steadily spinning down,
%and have no evidence of orbital periodicity to date (Israel et
%al. 1999a, Baykal et al. 1998, Mereghetti, Israel, \& Stella 1998,
%Vasisht \& Gotthelf 1997, Sugizaki et al. 1997, Israel et al. 1999b,
%Torii et al. 1998, Haberl et al. 1997).
APXs that have measured the spin-down rate (Israel et al. 1999,
Mereghetti, Israel, \& Stella 1998, Baykal et al. 1998, Vasisht \&
Gotthelf 1997, Haberl et al. 1997) can be considered to have magnetic
fields of $10^{14}$--$10^{15}$G.  Although there are some other models
for SGRs and AXPs, these objects are best understood within the
framework of magnetar (see, e.g., Thompson \& Duncan 1995, Kouveliotou
et al. 1998, Vasisht \& Gotthelf 1997). Therefore, from both
theoretical and observational results, it may be possible that a NS
has superstrong magnetic fields $\sim 10^{16}$G.

In this paper we investigate the effects of the magnetic dipole fields
of NSs on the frequency evolution of GWs from the inspiralling BNSs,
since the magnetic fields of order $\sim 10^{16}$G might cause about
one rotation error. Throughout the paper, we use the relation
$H=2\mu/R^3$ to connect the magnetic moment $\mu$ to the magnetic
field at the magnetic pole $H$, and $R=10^6$cm as the radius of a NS.
For later convenience, note that $\mu=1.4\times10^{9}(H/10^{16}{\rm
G})$ cm$^2$ in units of $G=c=1$.

\section{EQUATIONS OF MOTION}\label{sec:eom}

We consider a binary system of two compact bodies of masses, $m_1$ and
$m_2$, with magnetic dipole moments, ${\bm \mu}_1$ and ${\bm \mu}_2$,
respectively. Since we pay particular attention to the effects of
magnetic fields, we treat the orbital motion of BNSs in Newtonian
gravity. Although a spherical symmetry of the stellar configuration
is, in general, incompatible with the presence of magnetic fields
(Chandrasekhar 1981), we ignore quadrupole effects which are caused by
magnetic fields for a moment (see \S \ref{sec:discuss} and \S
\ref{sec:quadru}). We also neglect tidal effects which are expected
to be small until pre-merging phase of BNSs (Bildsten and Cutler
1992).

By eliminating the motion of the center of mass of BNSs and setting
the origin of the coordinate frame at the center of mass of BNSs, the
effective one-body equations of motion can be derived from a
Lagrangian,
\beqa
{\cal L}={{1}\over{2}} \eta m v^2 + {{\eta m^2}\over{r}}
+{\cal L}_{DD},
\label{eq:lag}
\eeqa
where
\beqa
{\cal L}_{DD}={\bm \mu}_1 \cdot {\bm H}_2
={{1}\over{r^3}}\left\{3(\bmn \cdot {\bm \mu}_1)
(\bmn \cdot {\bm \mu}_2) - {\bm \mu}_1 \cdot {\bm \mu}_2 \right\}.
\eeqa
Here $m=m_1+m_2$, $\eta=m_1 m_2/m^2$, $r=|{\bm x}|$, ${\bm x}={\bm
x}_1-{\bm x}_2$, $\bmn={\bm x}/r$, ${\bm v}=\dot {\bm x}$, and ${\bm
H}_2=\left\{3({\bmn \cdot \bm \mu}_2) {\bmn} - {\bm
\mu}_2\right\}/r^3$ is the magnetic field at ${\bm x}_1$ produced by
the magnetic moment ${\bm \mu}_2$. By using the Euler-Lagrange
equations, we obtain the equations of motion as
\beq
{\bm a}=-{{m}\over{r^2}}\bmn+{\bm a}_{DD},
\label{eq:eom}
\eeq
where ${\bm a}=\ddot {\bm x}$ and
\beq
{\bm a}_{DD}={{3}\over{\eta m r^4}}
\left\{({\bm \mu}_1 \cdot {\bm \mu}_2) \bmn + 
(\bmn \cdot {\bm \mu}_2) {\bm \mu}_1 + 
(\bmn \cdot {\bm \mu}_1) {\bm \mu}_2 -
5 (\bmn \cdot {\bm \mu}_1) (\bmn \cdot {\bm \mu}_2) \bmn \right\}.
\label{eq:aDD}
\eeq
{}From equation (\ref{eq:lag}), the energy of this system is given by
\beq
E={{1}\over{2}} \eta m v^2 - {{\eta m^2}\over{r}} + E_{DD},
\label{eq:energy}
\eeq
where
\beq
E_{DD}=-{{1}\over{r^3}}\left\{
3 (\bmn \cdot {\bm \mu}_1) (\bmn \cdot {\bm \mu}_2) - 
{\bm \mu}_1 \cdot {\bm \mu}_2 \right\}.
\label{eq:edd}
\eeq
The total angular momentum can be defined as
${\bm L}={\bm L}_N+{\bm S}$
where ${\bm L}_N=\eta m ({\bm x}{\bm \times}{\bm v})$ 
is the Newtonian orbital angular momentum and
${\bm S}={\bm S}_1+{\bm S}_2$ is the total spin angular momentum.
We can show explicitly $\dot E = \dot {\bm L} = 0$
with the equations of motion (\ref{eq:eom})
and the evolution equations of the spins,\footnote{
Note that there are no spin-orbit and spin-spin interactions
since we consider Newtonian gravity.
The spins are generated by the torque due to
the magnetic dipole-dipole interaction.}
\beq
\dot {\bm S}_p={\bm \mu}_p {\bm \times} {\bm H}_q
={{1}\over{r^3}}\left\{3(\bmn \cdot {\bm \mu}_q)({\bm \mu}_p
{\bm \times} \bmn) - {\bm \mu}_p {\bm \times} {\bm \mu}_q\right\}.
\quad (p,q=1,2)
\label{eq:spin}
\eeq
Assuming NSs as spherical compact bodies, the spin angular velocities
${\bm \Omega}_p$ are related to the spins ${\bm S}_p$ as ${\bm
S}_p=I_p {\bm \Omega}_p$ where $I_p$ is the principle moment of
inertia of the bodies. Since the magnetic moments evolve as $\dot
{\bm \mu_p}={\bm \Omega}_p {\bm \times} {\bm \mu}_p = {\bm S}_p {\bm
\times} {\bm \mu}_p/I_p$, the angular velocities of the magnetic
moments $\Omega_p$ will be of order $\Omega_p \sim (\mu_1 \mu_2/m R^2
r^3)^{1/2}$ from dimensional analysis with equation (\ref{eq:spin}).
Note also that the orbital angular velocity $w$ is of order 
$w \sim (m/r)^{3/2}/m$,
and the orbital inspiral rate $w_{ins}=(dE/dt)_{GW}/E$
is of order $w_{ins} \sim (m/r)^4/m$.

\section{GRAVITATIONAL WAVES AND ELECTROMAGNETIC WAVES}

As a first step, we use the quadrupole formula to derive the rate of
energy loss from a binary system due to GWs (e.g. Thorne 1980). The
symmetric, trace-free parts of the quadrupole moments of this system
are given by $\bari_{ij}=\eta m (x_i x_j - {{1}\over{3}} r^2
\delta_{ij})$. Taking third time derivatives of these quadrupole
moments, we obtain the energy loss rate from the quadrupole formula as
\beqa
\left({{dE}\over{dt}}\right)_{GW}
&=&-{{1}\over{5}}\left \langle 
{\dddot {\bari}}_{ij} {\dddot {\bari}}_{ij} \right\rangle \nonumber \\
&=&-{{8}\over{15}}{{\eta^2 m^4}\over{r^4}}
\Biggl\{12 v^2 - 11 \dot r^2 
\nonumber\\
&&+{{1}\over{\eta m^2 r^2}}\biggl[
6(-12 v^2 + 13 \dot r^2)({\bm \mu}_1 \cdot {\bm \mu}_2)
+12(21 v^2 - 34 \dot r^2)(\bmn \cdot {\bm \mu}_1)(\bmn \cdot {\bm \mu}_2)
\nonumber\\
&&-36({\bm v} \cdot {\bm \mu}_1)({\bm v} \cdot {\bm \mu}_2)
+87 \dot r \left\{(\bmn \cdot {\bm \mu}_1)({\bm v} \cdot {\bm \mu}_2)
+({\bm v} \cdot {\bm \mu}_1)(\bmn \cdot {\bm \mu}_2)\right\}
\biggr]\Biggr\},
\label{eq:lossGW}
\eeqa
where we have used the equations of motion (\ref{eq:eom}) and
assumed $\mu_1 \mu_2/m^2 r^2 \ll 1$.

On the other hand, electromagnetic (EM) waves are also emitted from
this binary system since the magnetic moments are moving. Using the
linearity in the EM fields, the radiation fields ${\bm B}^{rad}_{0}$
in equation (\ref{brad0}) for this binary system are given by
\beq
{\bm B}^{rad}_{0}={{1}\over{D}}(\bmd \cdot {\bm \mu}_{eff})
\left\{(\bmd \cdot \dot {\bm a}) \bmd - \dot {\bm a}\right\},
\label{eq:B0}
\eeq
where
\beq
{\bm \mu}_{eff}={{1}\over{m}}(m_2 {\bm \mu}_1 - m_1 {\bm \mu}_2).
\eeq
Therefore, since the radiated power is given by equation
(\ref{power}), the rate of energy loss due to the EM radiation is
calculated as
\beq
\left({{dE}\over{dt}}\right)_{EM}
=-{{2}\over{15}}{{m^2}\over{r^6}}
\left[2 \mu_{eff}^2 
\left\{v^2 - 6 \dot r (\bmn \cdot {\bm v}) + 9 \dot r^2 \right\}
- \left\{{\bm \mu}_{eff} \cdot ({\bm v} - 3 \dot r \bmn)\right\}^2 \right],
\label{eq:lossEM}
\eeq
where we have substituted ${\bm \mu}_{eff}$ into ${\bm \mu}$
in equation (\ref{power}).

Note that the assumption of the constant magnetic moments
in equations (\ref{eq:lossGW}), (\ref{eq:B0}) and (\ref{brad0}) is valid
since the angular velocities of the magnetic moments 
$\Omega_p\sim (\mu_1 \mu_2/m R^2 r^3)^{1/2}$
are much smaller than the orbital angular 
velocities $w \sim (m/r)^{3/2}/m$ when $(H_1 H_2)^{1/2} \ll 10^{18}$G.

\section{INSPIRAL OF CIRCULAR ORBITS}

For calculational simplicity we assume that the orbital motion of BNSs
has decayed to be circular apart from the adiabatic inspiral (Peters
\& Mathews 1963, Peters 1964). In general, circular orbit solutions of
equation (\ref{eq:eom}) do not exit unless the magnetic moments are
aligned perpendicular to the orbital plane. However, since the angular
velocities of the magnetic moments $\Omega_p \sim (\mu_1 \mu_2/m R^2
r^3)^{1/2}$ are much smaller than the orbital angular velocity $w \sim
(m/r)^{3/2}/m$ when $(H_1 H_2)^{1/2} \ll 10^{18}$G, we can regard the
magnetic moment vectors and $\bmL$ as time-independent ones over an
orbit where $\bmL$ is a unit vector orthogonal to the orbital plane.
Then, after taking an average of the magnetic term in the acceleration
(\ref{eq:aDD}), we can obtain orbits of constant separation, $\ddot
r=\dot r=\bmn \cdot {\bm v}=0$, $w=v/r$, ${\bm L}_{N}=\eta m r^2 w
\bmL$ (see similar discussions on spin precessions, Kidder, Will, \&
Wiseman 1993, Kidder 1995). From the equations of motion for circular
orbits, $\bmn \cdot {\bm a}=\ddot r-r w^2$, we can calculate the
orbital angular velocity as
\beq
w^2={{m}\over{r^3}}\left[1+{{3}\over{2 \eta m^4}}
\left({{m}\over{r}}\right)^2
\left\{{\bm \mu}_1 \cdot {\bm \mu}_2 - 3(\bmL \cdot {\bm \mu}_1)
(\bmL \cdot {\bm \mu}_2)\right\}\right],
\label{eq:w}
\eeq
where we have used the orbit-averaged relation,
\beq
\overline{(\bmn \cdot {\bm \mu}_1)(\bmn \cdot {\bm \mu}_2)}
={{1}\over{2}}
\left\{{\bm \mu}_1 \cdot {\bm \mu}_2 - (\bmL \cdot {\bm \mu}_1)
(\bmL \cdot {\bm \mu}_2)\right\}.
\eeq

The total energy and the energy loss rate for a circular orbit,
averaged over an orbit, can be obtained as
\beqa
-E&=&\eta {{m^2}\over{2 r}}
\left[1-{{1}\over{2 \eta m^4}} \left({{m}\over{r}}\right)^2
\left\{{\bm \mu}_1 \cdot {\bm \mu}_2 - 3 (\bmL \cdot {\bm \mu}_1)
(\bmL \cdot {\bm \mu}_2)\right\}\right],
\label{eq:energy2}
\\
{{dE}\over{dt}}&=&\left({{dE}\over{dt}}\right)_{GW}
+\left({{dE}\over{dt}}\right)_{EM}
\nonumber\\
&=&-{{32}\over{5}} \eta^2 \left({{m}\over{r}}\right)^{5}
\Biggl[1+{{9}\over{2 \eta m^4}} \left({{m}\over{r}}\right)^{2}
\left\{{\bm \mu}_1 \cdot {\bm \mu}_2 - 3 (\bmL \cdot {\bm \mu}_1)
(\bmL \cdot {\bm \mu}_2)\right\}
\nonumber\\
&&+{{1}\over{96 \eta^2 m^4}} \left({{m}\over{r}}\right)^{2}
\left\{3 \mu_{eff}^2 + ({\bm \mu}_{eff} \cdot \bmL)^2 \right\}
\Biggr],
\label{eq:loss}
\eeqa
by using equations (\ref{eq:energy}), (\ref{eq:lossGW}),
(\ref{eq:lossEM}) and (\ref{eq:w}).

Combining equations (\ref{eq:w}), (\ref{eq:energy2}) and (\ref{eq:loss}),
we can express the change rate of the orbital angular velocity $\dot w$
as a function of $w$,
\beq
\dot w = {{96}\over{5}} \eta m^{5/3} w^{11/3}
\left\{1+\sigma_{mag}(m w)^{4/3}\right\},
\label{eq:wdot}
\eeq
where 
\beq
\sigma_{mag}={{5}\over{\eta m^4}}
\left\{{\bm \mu}_1 \cdot {\bm \mu}_2 - 3 (\bmL \cdot {\bm \mu}_1)
(\bmL \cdot {\bm \mu}_2)\right\}
+{{1}\over{96 \eta^2 m^4}}
\left\{3 \mu_{eff}^2 + ({\bm \mu}_{eff} \cdot \bmL)^2 \right\}.
\label{eq:sigmag}
\eeq
By using equation (\ref{eq:wdot}), we calculate the accumulated number
of GW cycles $N=\int (f/\dot f) df$, where $f=w/\pi$ is the frequency
of the quadrupolar waves. For calculational simplicity, we assume
that the two magnetic moments are aligned parallel to the orbital axis
of the binary system. In this case, $\sigma_{mag}$ becomes a constant
value. Then, the accumulated number is integrated as
\beq
N=N_{grav} +N_{mag},
\eeq
where $N_{grav}$ and $N_{mag}$ denote the contributions from the
Newtonian gravity term and the magnetic term
respectively, and are expressed as
\beqa
N_{grav} &=&-\left.{{1}\over{32\pi \eta}}(\pi m f)^{-5/3}
\right|^{f_{max}}_{f_{min}}, \\
N_{mag} &=&\left. {5 \over 32\pi \eta} \sigma_{mag} (\pi m
f)^{-1/3}\right|^{f_{max}}_{f_{min}}.
\eeqa
Here $f_{max}$ is the exit frequency and $f_{min}$ is the
entering one of the detector's bandwidth.

\section{DISCUSSION}\label{sec:discuss}

Using 10Hz as the entering frequency and 1000Hz as the exit one, we
obtain the contribution to the total number of GW cycles from the
magnetic term as
\beq
N_{mag}=-5.9\times 10^{-3}
\left({{H_1}\over{10^{16}{\rm G}}}\right)
\left({{H_2}\over{10^{16}{\rm G}}}\right)
\left({{m}\over{2\times 1.4 M_{\odot}}}\right)^{-13/3},
%\left({{R}\over{10^{6}{\rm cm}}}\right)^{6},
\eeq
where we assume ${\bm \mu}_1 \cdot {\bm \mu}_2 < 0$,
${\bm \mu}_1 \parallel {\bmL}$, ${\bm \mu}_2 \parallel {\bmL}$,\footnote{
This set of configurations, ${\bm \mu}_1 \cdot {\bm \mu}_2 < 0$, ${\bm
\mu}_1 \parallel {\bmL}$ and ${\bm \mu}_2 \parallel {\bmL}$, is the
most stable one, i.e., the EM interaction energy in equation
(\ref{eq:edd}) becomes the minimum value. However, note that the
alignment rate of the magnetic moments $w_{ali}$ is smaller than the
orbital inspiral rate $w_{ins} \sim (m/r)^4/m$ when $H \siml 10^{16}
(r/10^{11}{\rm cm})^{-1/4}$ G, since the alignment rate $w_{ali}$ can
be estimated as $w_{ali} \sim ({\rm energy\ loss\ rate\ due\ to\
magnetic\ dipole\ radiation})/E_{DD} \sim (\mu \Omega_p^2)/(\mu^2/r^3)
\sim \mu^4/r^3 m^2 R^4$ when $\mu_1 \sim \mu_2 \sim \mu$.
}
and $m_1=m_2$. 
Note
that the contribution of the EM radiation reaction (the second term in
equation (\ref{eq:sigmag})) is much less than that of the
dipole-dipole interaction (the first term in equation
(\ref{eq:sigmag})). The maximum magnetic field allowed by the scalar
virial theorem is $\sim 10^{18}$G for NSs (Chandrasekhar 1981, see
also Bocquet et al. 1995). If NSs in the inspiralling BNSs have such
magnetic fields $10^{17} {\rm G} \siml H \siml 10^{18} {\rm G}$, the
effects of the magnetic fields can change more than one cycle in the
accumulated cycles. However, if we consider that the maximum value of
the observed fields $\sim 10^{16}$ G is the upper limit for the
magnetic fields of NSs, the magnetic term will make negligible
contributions to the accumulated phase, contrary to the crude estimate
in \S \ref{sec:intro} and equation (\ref{eq:crude}). Consequently,
the magnetic fields of NSs will not present difficulties for the
detection of GWs from BNSs, if the upper limit for the magnetic fields
of NSs is less than $\sim 10^{16}$G.

The magnetic term in equation (\ref{eq:wdot}) has the same dependence
on the angular velocity $w$ as 2PN terms. We can see this dependence
from the 2PN expression for the frequency sweep (Blanchet et
al. 1995),
\beqa
\dot w &=& {{96}\over{5}} \eta m^{5/3} w^{11/3}
\left\{1-\left({{743}\over{336}}+{{11}\over{4}}\eta\right)(m w)^{2/3}
+(4\pi-\beta)(m w)\right.
\nonumber\\
&&+ \left.\left({{34103}\over{18144}}+{{13661}\over{2016}}\eta
+{{59}\over{18}}\eta^2+\sigma \right)(m w)^{4/3}\right\},
\label{eq:2PN}
\eeqa
where the spin-orbit parameter is $\beta={{1}\over{12}}\Sigma_p(113
m_p^2/m^2+75\eta) \hat{\bm L}\cdot{\bm \chi}_p$, the spin-spin
parameter is $\sigma=({{\eta}/{48}})(-247 {\bm \chi}_1 \cdot {\bm
\chi}_2 +721\hat{\bm L}\cdot{\bm \chi}_1 \hat{\bm L}\cdot{\bm
\chi}_2)$ and ${\bm \chi}_p={\bm S}_p/m_p^2$. As we can see from
equations (\ref{eq:wdot}) and (\ref{eq:2PN}), we cannot distinguish
the magnetic term $\sigma_{mag}$ from the spin-spin term $\sigma$ at
2PN order. Only the sum of the magnetic term and the spin-spin term
can be deduced from the frequency evolution of 2PN order. This
degeneracy might be broken by examining the wave-form modulations
caused by the spin-induced precession of the orbit (Apostolatos et
al. 1994, Kidder 1995).
However, the analysis of Poisson \& Will (1995) shows that the
measurement error on the spin-spin term is $\Delta \sigma \sim 17.3$
assuming that the SNR is 10, $\beta=0$,
$\sigma=0$ and there is no prior information.\footnote{ Because of
some simplifying assumptions in Poisson \& Will (1995), the true
measurement error is still uncertain. However, this will not
affect the discussion since the true measurement error will be within
a factor of the order of 2 (Balasubramanian, Sathyaprakash, \&
Dhurandhar 1996, Balasubramanian \& Dhurandhar 1998, Nicholson \&
Vecchio 1998).
}
On the other hand, 
the contribution of the magnetic term is estimated as
\beq
\sigma_{mag}=2.9 \times 10^{-3}
\left({{H_1}\over{10^{16}{\rm G}}}\right)
\left({{H_2}\over{10^{16}{\rm G}}}\right)
\left({{m}\over{2\times 1.4 M_{\odot}}}\right)^{-4},
%\left({{R}\over{10^{6}{\rm cm}}}\right)^{6},
\eeq
where we assume ${\bm \mu}_1 \cdot {\bm \mu}_2 < 0$,
${\bm \mu}_1 \parallel {\bmL}$, ${\bm \mu}_2 \parallel {\bmL}$
and $m_1=m_2$.
We can see from this equation that the contribution of the magnetic
fields of NSs is much smaller than the measurement error on the
spin-spin term. Therefore, the effects of the magnetic fields of NSs
are also negligible for parameter estimation with moderate SNR if we
consider that the maximum value of the observed fields $\sim 10^{16}$G
is the upper limit for the magnetic fields of NSs.

Conservatively, there are some other reasons to consider that the
magnetic fields of inspiralling NSs are smaller than $\sim 10^{16}$G.
There are three observed BNSs in our Galaxy and nearby globular
cluster that will merge in less than a Hubble time: PSR B1913+16
(Taylor \& Weisberg 1989), PSR B1534+12 (Wolszczan 1991, Stairs et
al. 1998) and PSR B2127+11C (Prince et al. 1991). From the spin-down
rate, the magnetic field strength is estimated as of order $10^{10}$G
for all pulsars in these binary systems. If such a value is typical of
the magnetic field strength,\footnote{ However, note that no radio
pulsars have magnetic fields above $\sim 10^{14}$G, and hence there
may be a selection bias (Baring \& Harding 1998).} the magnetic terms
will be negligible. Moreover, if the decay time of the magnetic
fields is shorter than the coalescence time (Heyl and Kulkarni 1998,
Thompson and Duncan 1996, Goldreich \& Reisenegger 1992, Shalybkov and
Urpin 1995), of course, the magnetic fields will not be concerned,
although the magnetic field evolution of isolated NSs is an unresolved issue. 
It may be also difficult for BNSs including magnetar to be formed because
of a large recoil velocity (Duncan \& Thompson 1992).

In this paper, we regard NSs as spherical compact bodies. However,
when we consider NSs as extended bodies, we have to take into account
of the quadrupole effects induced by magnetic fields. The magnetic
fields are a source of non-hydrostatic stress in the interiors of
NSs. A magnetic dipole moment $\mu$ would give rise to moment
differences of order
\beq
\epsilon:={{I_c-I_a}\over{I_a}}\sim {{R^4 H^2}\over{m^2}},
\label{eq:ellip}
\eeq
where $I_c$ and $I_a$ refer to the moments of inertia about the dipole
axis and about an axis in the magnetic equator. Then, the
gravitational potential $\eta m^2/r$ between NSs is modified by an
amount of order $m I_a \epsilon/r^3 \sim \mu^2/r^3$. This is the same
order as the EM interaction term in equation (\ref{eq:lag}) when
$\mu_1 \sim \mu_2$. Therefore, an accurate ellipticity $\epsilon$ in
equation (\ref{eq:ellip}) is needed to determine the magnetic effects
within a factor. (Bocquet et al. 1995, Bonazzola \& Gourgoulhon
1996, Konno, Obata, \& Kojima 1999). 
Note that there are such quadrupole effects even if only one
companion has magnetic moment.

Although the effects of the magnetic fields of NSs will be negligible
for observations of GWs, they might be concerned with gamma ray bursts
(GRBs). The BNS merger is one of models of GRB sources. (see, e.g.,
Piran 1999 for a review). If a NS in the binary system has strong
magnetic fields, the total energy emitted by EM waves until
coalescence can be estimated from equations (\ref{eq:w}),
(\ref{eq:loss}) and (\ref{eq:wdot}) as $-\int (dE/dt)_{EM} dt \sim
\pi^2 \mu_{eff}^2 (f_{max}^2 - f_{min}^2) /144 \eta m \sim 10^{46}
(H/10^{16} {\rm G})^2 (f_{max}/10^3{\rm Hz})^2$ ergs. This energy will
be radiated at very low frequency $\sim 10^{3}$ Hz, which is difficult
to be observed by the present radio telescope. Furthermore such
radiation cannot propagate a plasma if an electron density is larger
than $\sim 0.01 {\rm cm}^{-3}$ since the plasma frequency is larger
than the radiation frequency (e.g. Spitzer 1962). However, this energy
may be converted to the thermal energy of the surrounding plasma
efficiently if the electron density $n_e$ is sufficiently high since
the electron-electron relaxation time is about $t_{rel}\sim 1
(n_e/10^{11}{\rm cm}^{-3})^{-1} (k T_e/2 {\rm keV})^{3/2}$ s where
$T_e$ is the electron temperature (e.g. Spitzer 1962). Then, this
thermal radiation might explain the precursory X-ray emission $\sim
10$ s before the onset of the GRB observed by the Ginga satellite, in
which the total energy of the X-ray precursor emission is estimated to be
about $\sim 10^{46} (d/100 {\rm Mpc})^{2}$ ergs (Murakami et
al. 1991).
Even though the strong magnetic fields are not relevant to the X-ray
precursor in GRB, the EM radiation can be the EM signature of the 
coalescing BNSs.
Therefore it is an interesting future problem to investigate
the conversion of the low frequency EM radiation
to the higher frequency one.

\acknowledgments
We would like to thank H. Sato and T. Nakamura for continuous
encouragement and useful discussions. 
We are also grateful to T. Tanaka, R. Nishi, K. Nakao,
T. Harada and K. Omukai for useful discussions.
This work was supported in part by
Grant-in-Aid for Scientific Research Fellowship (No.9627: KI) and
(No.9402: KT) of the Japanese Ministry of Education,
Science, Sports and Culture.

%%%%%%%% appendix %%%%%%%%%%%%%%%%%%%%%%
\appendix

\section{ELECTROMAGNETIC RADIATION FROM A MOVING MAGNETIC MOMENT}

In this section we review the radiation from a moving magnetic dipole
moment. We consider a particle with only a magnetic dipole moment
${\bm \mu}'$ in its rest frame $K'$. A moving magnetic dipole moment with
velocity ${\bm v}=\dot {\bm x}$ relative to an observer frame $K$ also
has an associated electric dipole moment. The apparent electric
dipole moment is
\beq
{\bm p}={\bm v} {\bm \times} {\bm \mu},
\label{edip}
\eeq
where ${\bm \mu}={\bm \mu}'- {{\gamma}\over{\gamma+1}}({\bm v} \cdot
{\bm \mu}'){\bm v}$ is the magnetic moment observed in K and
$\gamma=(1-v^2)^{-1/2}$ (Jackson 1998). Therefore, in the observer
frame $K$ the magnetization density ${\bm M}$ and electric
polarization density ${\bm P}$ are given by
\beqa
{\bm M}(t,{\bm z})&=&{\bm \mu}(t) \delta[{\bm z}-{\bm x}(t)],
\nonumber\\
{\bm P}(t,{\bm z})&=&{\bm p}(t) \delta [{\bm z}-{\bm x}(t)],
\eeqa
where ${\bm p}$ is given by equation (\ref{edip}). Recalling that the
moving magnetic moment is equivalent to a current ${\bm J}=\nabla {\bm
\times} {\bm M} + \dot {\bm P}$ (Jackson 1998), we can calculate the
electric and magnetic fields. For our purpose, it is sufficient to
obtain the radiative parts which fall off as the inverse of the
distance $D^{-1}$. The radiation field ${\bm B}^{rad}$ of this
moving magnetic moment is given by (e.g. Heras 1994)
\beqa
{\bm B}^{rad}(t,{\bm z})
&=&{{3 {\bmd} {\bm \times} ({\bmd} {\bm \times} {\bm \mu}
- {\bm p})({\bmd} \cdot {\bm a})^{2}}\over
{D W^5}}
+{{3 {\bmd} {\bm \times} ({\bmd} {\bm \times} \dot{\bm \mu}
- \dot {\bm p})({\bmd} \cdot {\bm a})}\over
{D W^4}}
\nonumber\\
&&+{{{\bmd} {\bm \times}({\bmd} {\bm \times} {\bm \mu}
-{\bm p})({\bmd} \cdot \dot {\bm a})}\over
{D W^4}}
+{{{\bmd}{\bm \times}({\bmd}{\bm \times} \ddot {\bm \mu}
- \ddot {\bm p})}\over
{D W^3}},
\label{Brad}
\eeqa
where ${\bm D}(t)={\bm z}-{\bm x}(t)$, ${\bmd}={\bm D}(t)/|{\bm
D}(t)|={\bm D}(t)/D$, $W=1-{\bm v} \cdot {\bmd}$, ${\bm a}=\dot {\bm
v}$ and the right-hand side of this equation is evaluated at the
retarded time $t'$, i.e., $t'+D(t')=t$. When we assume that the
magnetic dipole moment vector ${\bm \mu}(t)$ is a constant one, the
above equation (\ref{Brad}) can be calculated as
\beqa
{\bm B}_{0}^{rad}&=&{{1}\over{D}}
\left\{{\bmd} {\bm \times} ({\bmd} {\bm \times} {\bm \mu})
({\bmd} \cdot \dot {\bm a}) - {\bmd} {\bm \times} \ddot {\bm p}\right\}
\nonumber\\
&=&{{1}\over{D}}
({\bmd} \cdot {\bm \mu}) 
\left\{({\bmd} \cdot \dot{\bm a}) {\bmd} - \dot {\bm a}\right\},
\label{brad0}
\eeqa
up to the leading order term of ${\bm v}$ and $|{\bm x}|/D$. Note
that the radiation field ${\bm B}_0^{rad}$ becomes the same equation
as (\ref{brad0}) up to leading order even if we use ${\bm \mu}'$
instead of ${\bm \mu}$. The power radiated per unit solid angle is
given by (e.g. Landau \& Lifshitz 1975)
\beq
{{dP_0}\over{d\Omega}}={{1}\over{4\pi}} 
\left({B_0^{rad}}\right)^2 D^2
={{1}\over{4\pi}}({\bmd} \cdot {\bm \mu})^2
\left\{|\dot {\bm a}|^2 - ({\bmd} \cdot \dot {\bm a})^2\right\}.
\label{dpower}
\eeq
The total instantaneous power is obtained by integrating 
equation (\ref{dpower}) over all solid angle as
\beq
P_0={{2}\over{15}}\left\{2 \mu^2 |\dot {\bm a}|^2 - 
({\bm \mu} \cdot \dot {\bm a})^2\right\}.
\label{power}
\eeq

\section{THE CONTRIBUTION OF ELECTROMAGNETIC FIELDS TO THE MASS}
\label{sec:quadru}

Thus far we have ignored the contribution of the EM fields to the mass
of a compact body in a binary system. In this section, we estimate the
correction to the mass by the EM fields. 
We have implicitly defined the mass $m$ as
that of the isolated spherical body, i.e., the orbital separation of the
binary system is infinity. Therefore, the mass $m$ includes the
self-energy of the EM fields. The self-energy $m_{mag}$ of the
magnetic fields outside the compact body with a magnetic dipole moment
${\bm \mu}$ is obtained by
\beq
m_{mag}=\int {{H^2}\over{8\pi}} d^3x'
={{1}\over{8\pi}}\int {{3 \mu^2 \cos^2\theta' + \mu^2}\over{r'^6}} d^3x'
=\mu^2 \left[-{{1}\over{3 r'^3}}\right]^{\infty}_{R}
={{\mu^2}\over{3 R^3}},
\label{eq:self}
\eeq
where $R$ is the radius of the compact body.
(The effects of the magnetic fields inside the compact body are discussed
in \S \ref{sec:discuss}.)

Next, we consider the gravitational interaction between the above
compact bodies with magnetic moments. Here we note that 
at a finite separation the
gravitational field of a mass with magnetic dipole moment ${\bm \mu}$
is different from that of a point mass without magnetic dipole moment
even if the masses are the same value
since EM fields have an extent. Therefore, the mass $m$ in the
gravitational potential term $\eta m^2/r$ in equation (\ref{eq:lag})
suffers a small correction. This correction can be estimated by
evaluating the ``gravitational potential'' produced by the energy of
the EM fields of the magnetic dipole moment $\bm \mu$. The
``gravitational potential'' $\phi$ at ${\bm x}$ is obtained by
\beqa
\phi({\bm x})=-\int {{H^2}\over{8\pi |{\bm x}-{\bm x}'|}} d^3x'
=-{{\mu^2}\over{8\pi}} \int {{3 \cos^2 \theta' + 1}
\over{r'^6 |{\bm x}-{\bm x}'|}} d^3x'.
\label{eq:potential}
\eeqa
Here we expand $1/|{\bm x}-{\bm x}'|$ by the spherical harmonics 
$Y_{lm}(\theta, \varphi)$,
\beq
{{1}\over{|{\bm x}-{\bm x}'|}}
=4\pi \sum^{\infty}_{l=0} \sum^{l}_{m=-l}{{1}\over{2l+1}}
{{r^l_{<}}\over{r^{l+1}_{>}}} Y_{lm}^{*}(\theta', \varphi')
Y_{lm}(\theta, \varphi),
\label{eq:expand}
\eeq
where $r_{<} (r_{>})$ is the smaller (larger) one of $|{\bm x}|$ and
$|{\bm x}'|$.
Noting the orthonormality of the spherical harmonics,
$\int Y_{l'm'}^{*}(\theta, \varphi) Y_{lm}(\theta, \varphi) d\Omega
=\delta_{l'l} \delta_{m'm}$,
and a relation, 
$3 \cos^2 \theta' + 1 = 2 \sqrt{4\pi/5} Y_{20}(\theta', \varphi')
 + 2 \sqrt{4\pi} Y_{00}(\theta', \varphi')$,
the equation (\ref{eq:potential}) is calculated as
\beqa
\phi({\bm x}) =
%-{{2}\over{5}} \sqrt{{{\pi}\over{5}}} \mu^2 
%Y_{20}^{*}(\theta', \varphi') \left(\int^{r'}_{R} {{1}\over{x^2 x'^3}} dx
%+ \int^{\infty}_{r'} {{x'^2}\over{x^7}} dx\right)
%\nonumber\\
%&-& \sqrt{4\pi} \mu^2 Y_{00}^{*}(\theta', \varphi')
%\left(\int^{x'}_{R} {{1}\over{x^4 x'}} dx
%+ \int^{\infty}_{x'} {{1}\over{x^5}} dx\right)
%\nonumber\\
- {{1}\over{r}} \left[{{\mu^2}\over{3 R^3}}
+ {{\mu^2 (3 \cos \theta -1)}\over{10 R r^2}}
- {{\mu^2 \cos^2 \theta}\over{4 r^3}} \right].
\eeqa
The first term in the bracket on the right-hand side of the above
equation comes from the total self-energy of the EM fields in equation
(\ref{eq:self}), and the last two terms are the corrections due to the
extent of the EM fields. The order of the correction to the mass is
estimated as ${{\delta m}/{m}} \sim (\mu^2/R r^2)/m \sim {{\mu^{2}}/{m
R r^2}}$. On the other hand, the correction due to the dipole-dipole
interaction is of order $\mu_1 \mu_2/m^2 r^2$ from equation
(\ref{eq:wdot}). Therefore, the correction due to the extent of the
magnetic fields are smaller than that due to the dipole-dipole
interaction when $\mu_1\sim \mu_2$. We have also confirmed that the
contribution of the EM fields to the quadrupole moments $\bari_{ij}$
is of order ${{\mu^{2}}/{m R r^2}}$.

\end{document}